# The nanostructural origin of the ac conductance in dielectric granular metals: the case study of $Co_{20}(ZrO_2)_{80}$

Z Konstantinović, M García del Muro, X Batlle, and A Labarta

Departament de Física Fonamental and Institut de Nanociència i Nanotecnologia,

(IN2UB), Universitat de Barcelona, Martí i Franquès 1, 08028-Barcelona, Spain

M Varela

Departament de Física Aplicada i Òptica and Institut de Nanociència i Nanotecnologia,

(IN2UB), Universitat de Barcelona, Martí i Franquès 1, 08028-Barcelona, Spain

(Dated:)

We show which is the nanostructure required in granular $Co_{20}(ZrO_2)_{80}$ thin films to produce an ac response such as the one that is *universally* observed in a very wide variety of dielectric materials. A bimodal size distribution of Co particles yields randomly competing conductance channels which allow both thermally assisted tunneling through small particles and capacitive conductance among larger particles that are further apart. A model consisting on a simple cubic random resistance-capacitor network describes quantitatively the experimental results as functions of temperature and frequency, and enables the determination of the microscopic parameters controlling the ac response of the samples.

Granular films are very promising nanostructured materials[1] with several technological applications including high-coercive films for information storage[2], high-permeability high-resistivity films for shielding and bit writing at high frequencies[3], and giant magnetoresistance for read heads and magnetic sensors[4]. However, their practical use in electro-optical devices requires a deeper knowledge of the microscopic mechanisms governing the low-frequency dielectric response, such as the occurrence of a dielectric-loss peak at low frequencies reported in granular Co-$ZrO_2$ thin films[5,6]. In fact, the latter ac response is shared by a wide variety of disordered solids (the so-called *universal dielectric response*)[7] which remains a topic of active experimental and theoretical research[8,9,10,11]. The common feature for all of these systems is the existence of disordered electrical paths which are intimately related to their microstructure. In particular, the observed ac behavior in granular Co-$ZrO_2$ was attributed to the competition between two channels of interparticle conductance at the nanoscale: thermally assisted tunneling and capacitance[5,6]. Actually, controlling and fully understanding the interplay between the nanostructure of granular metals and their electrical properties is a crucial issue to develop applications of these materials.

In this letter, we report on the nanostructure that is required for the observation of *universal* ac response in granular $Co_{20}(ZrO_2)_{80}$, which has been chosen as a model system because of both the appropriate metal content and low degree of metal-oxide interdiffusion. The microstructure is extremely clean displaying sharp interfaces between the crystalline metallic Co particles and the amorphous $ZrO_2$ matrix. High-resolution transmission microscopy (HRTEM) clearly shows a bimodal size distribution of metallic particles in samples exhibiting the *universal* ac response. The first component of this distribution corresponds to particles of a few nanometers in size, close enough to enable electrical tunneling between them, while the second one is centered at larger sizes corresponding to particles further apart that enable parallel conductance channels through inter-particle capacitance. It is also worthwhile to emphasize that this observed nanostructure is well in accordance with the former hypothesis of Abeles et al., about the *homogeneity* of the metallic phase in granular solids, which implies that the ratio between the inter-particle distance and particle diameter is roughly constant for particles of similar sizes. Besides, the observed size heterogeneity in $Co_{20}(ZrO_2)_{80}$ may also give insight in other open questions in granular materials such as the occurrence of high-order tunneling processes between large particles mediated by small particles proposed by Mitani et al.[12] in order to explain the strong increase of the tunneling magnetoresistance observed at low temperatures[13].

$Co_x(ZrO_2)_{100-x}$ granular films were prepared by laser ablation with compositions ranging from metallic to dielectric regimes[5,6,13]. Low-frequency absorption phenomena were observed in the dielectric regime ($0.18<x<0.31$) close to the percolation threshold. In this letter, we focus on granular $Co_{20}(ZrO_2)_{80}$, for which the existence of particle-size heterogeneity is clearly demonstrated from both structural and ac transport properties. HRTEM micrographs show a bimodal distribution of Co particles: i) small ones which are very close and ii) large ones further apart one to each other (Fig.1), both of them being randomly distributed throughout the sample. The presence of lattice fringes in the metal grains indicates their good crystallinity (left-hand inset to Fig.1) (notice their absence in the amorphous matrix). Co particles are mostly spherical and show sharp interfaces with the matrix without detectable interdiffusion boundaries. The histogram of particle sizes (right-hand inset to Fig.1) obtained from HRTEM micrographs can be fitted by the superimposition of two log-normal distributions corresponding to 83% of small particles and 17% of large ones. These distributions yield average particle diameters of $D_1$=5.7±0.2 nm and $D_2$=17.0±0.5 nm, with $\delta_1$=0.24±0.02 and $\delta_2$=0.20±0.02 being the values of the distribution widths, respectively. The average inter-particle distance, *d*, is about 5 nm for the first log-normal distribution and 15 nm for the second one, in agreement with the Abeles hypothesis of uniform composition in granular media[14] ($d/D \approx$ constant).

Ac electrical measurements were carried out with a four probe technique using a lock-in amplifier and a constant current source in the frequency range of 10-6000 Hz. The dc resistance was extrapolated from measurements at 12 Hz. The temperature dependence of the dc resistance, *R*, for $Co_{20}(ZrO_2)_{80}$ can be accounted for by $R \sim \exp(2\sqrt{B/k_BT})$ with *B*=22 meV, which indicates the occurrence of thermally assisted tunneling between particles. Tunneling paths going across small particles separated by no more than a few Angstroms and covering the whole sample can be easily sketched in the HRTEM micrograph in Fig.1. The impedance of the sample increases with decreasing temperature for all frequencies since thermal activation is necessary to overcome the charging energy of the particles (Coulomb gap) in order that tunneling processes take place. On the other hand, the impedance of the sample decreases with increasing frequency at all temperatures showing that capacitive paths are also present. This decrease with frequency is more abrupt as the temperature decreases, indicating that capacitive contribution to the ac conduction is more significant at low temperatures where the tunneling processes are reduced because of the low amount of charge carriers.

Deeper insight in the nature of the electrical paths existing throughout the sample can be gained by analyzing the phase $\varphi$ of the voltage drop across the sample as a function of the frequency $v$ and temperature $T$ (Fig. 2). Fig.2(a) shows the results for the surface $\varphi(v,T)$. At room temperature and low frequencies close to the dc limit, $\varphi$ is slightly negative, indicating that conduction is mostly due to electrons tunneling between nanoparticles and flowing through resistive paths. With increasing frequency, $\varphi$ diminishes as the current due to interparticle capacitance becomes progressively more important as compared to the tunneling contribution. When the capacitive and tunneling currents become comparable, a minimum develops at a frequency $v_{min}(T)$ which is temperature-dependent (see $\varphi(v)$ at 290 K in Fig.2(b)). For $v>v_{min}(T)$, $\varphi$ becomes less negative slowly approaching a maximum value due to the introduction of shortcuts of capacitive nature between particles at higher distances, which establish new paths for the tunneling current, in such a way that the real part of the electrical conductivity of the sample increases. As the temperature is lowered, thermally-assisted tunneling processes are less effective, increasing the capacitive nature of the total current. Consequently, $v_{min}(T)$ shifts towards smaller frequencies becoming unobservable below about 100 K (see $\varphi(v)$ at 100 K in Fig.2(b)). At $T \leq 100$ K tunneling processes are strongly reduced so the capacitive current dominates at moderate values of the frequency. Accordingly, $\varphi(v)$ shows a broad peak and decreases monotonously as $v$ is further increased (see $\varphi(v)$ at 50 K in Fig.2(b) and inset to Fig.2(b)).

The observed ac properties can be modeled by a random resistor-capacitor R-C network[15]. For simplicity we have chosen a cubic lattice in which each site is linked with its six nearest neighbors by randomly distributed connections of capacitive/resistive nature. In particular, the studied R-C network has 10*10*10 sites, size for which finite-size effects are virtually undetectable. In this lattice, a fraction $x_r$ of the connections is a parallel arrangement of a resistance $R_t$ and a low capacitance $C_p$ representing tunneling paths between small particles, and a fraction (1-$x_r$) is a high capacitance $C_p{'}$ representing capacitive paths through larger particles which are further apart. The phase of the voltage drop between two parallel plates connected to the network is calculated with the WINSPICE software[16], with $x_r$, $R_t$, $C_p$ and $C_p{'}$ as adjustable parameters. For 17 nm particles with an average separation of 15 nm, $C_p{'}$ is about $2 \cdot 10^{-18}$ F[17]. Assuming that $C_p{'}$ is temperature independent, an estimation of the rest of the parameters is obtained by fitting experimental data for $\varphi(v,T)$. A comparison between experimental data and the corresponding results for the R-C network at three temperatures is shown in Fig. 2(b): the fitted curves reproduce the overall behavior of $\varphi(v)$ in the whole range of measured $v$ and $T$. The resistive fraction $x_r$ is essentially fixed by the negative value of the phase at $v_{min}(T)$ and it is found to be nearly temperature-independent ($x_r$=0.835 at 290 K and $x_r$=0.80 at T=50 K). It is worth noting that this value is in very good agreement with the percentage of small particles (83%) estimated from HRTEM analysis, so $x_r$ can be associated to the fraction of small particles through which tunneling takes place, while (1-$x_r$) corresponds to the fraction of larger particles connected only by capacitance. The temperature-dependence of the fitted values of $R_t$ and $C_p$ is plotted in Fig. 3. It is noticeable that $R_t(T)$ (squares in Fig. 3)- which can be associated with the average tunneling resistance between neighboring particles - scales well (by simply multiplying by an arbitrary factor) with the temperature dependence of the dc resistance of the sample (solid line in Fig. 3), showing the expected behavior for thermally-assisted tunneling in the dielectric regime of a granular metal[14,5,6]. In contrast, $C_p$ is almost constant (triangles in the inset to Fig. 3). The large values of the ratio $C_p'/C_p=10^3$-$10^4$ suggest that $C_p$ cannot be directly associated with the average value of the capacitance between two of the smallest particles facing one to each other, which are shown in the HRTEM image. In fact, it is reasonable to assume that the actual amount of very small particles (of a few nm in size) may be larger than what HRTEM image indicates, since they could not be well resolved in a 0.5x0.5 µm image. If this was the case, each resistive branch of the sc network would correspond to the average of a small region containing certain amount of particles linked by tunneling paths among them. Therefore, $C_p$ may represent the resulting capacitance of a series-parallel network of the elementary capacitors connecting these particles, thus giving rise to a much more reduced value than that corresponding to the capacitance between two small particles. Finally, the orders of magnitude of the fitted values for $R_t \sim 10^{15}$-$10^{16}$ Ω are reasonable as compared with those values estimated by using experimental data on tunneling junctions[18] for particles with a separation of 1 nm and an effective tunneling surface of 1 nm$^2$ ($R_t \sim 10^{16}$-$10^{14}$ Ω as the temperature varies from 30 to 290 K).

In conclusion, we show the kind of particle-size distribution that is required to produce an *universal* ac behavior in granular $Co_{20}(ZrO_2)_{80}$ thin films. This behavior occurs at lower frequencies as compared with other disordered dielectric materials. The ac response is a direct consequence of the random competition between two parallel conductance channels: tunneling through small particles and capacitive conduction through larger particles that are further apart. This nanostructure is confirmed by HRTEM micrographs. The bimodal size distribution found in $Co_{20}(ZrO_2)_{80}$ is inherent to many other granular solids and might give a further insight in some other open questions in these materials, such as the strong increase of the tunneling magnetoresistance observed at low temperatures in clean samples[4,13].

We would like to thank the staff of the scientific and technical facilities of the University of Barcelona. Financial support of the Spanish CICYT (MAT2006-03999) and Catalan DURSI (2005SGR00969) are gratefully recognized. ZK thanks the Spanish MEC for the financial support through the Juan de la Cierva program.

**FIGURE CAPTIONS:**

**FIG 1.** HRTEM micrograph of granular $Co_{20}(ZrO_2)_{80}$. Left-hand side inset shows the magnification of one Co particle. Right-hand side inset shows the distribution of particle sizes obtained by analyzing HRTEM micrographs.

**FIG. 2.** (color online) (a) Experimental results for the $\varphi(v,T)$ surface. (b) Experimental results for $\varphi(v)$ at 290 K (squares), 100 K (circles) and 50 K (triangles) compared with the fits to a random R-C network (solid lines).

**FIG. 3.** Temperature dependence of the total dc resistance of the sample (solid line) as compared to $R_t(T)$ (squares) obtained from the fit of $\varphi(v,T)$ to a random R-C network. The inset shows the temperature dependence of $C_p$ (triangles) and $C_p`$ (circles).

Fig.1.

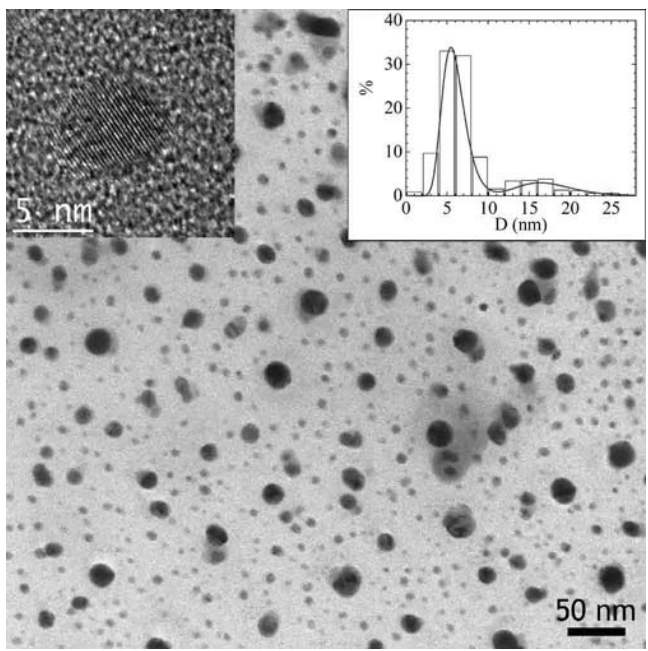

Fig.2

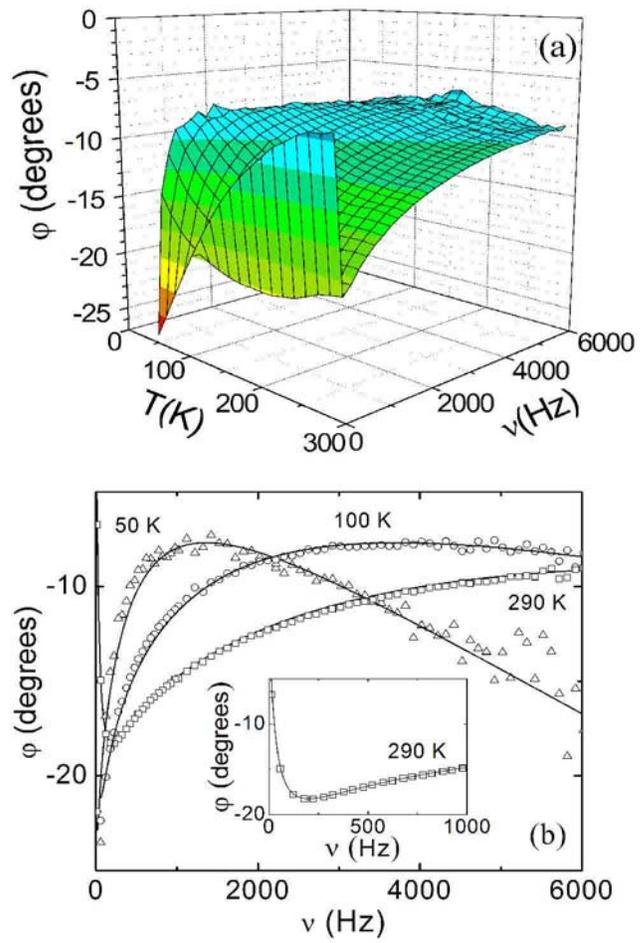

Fig.3

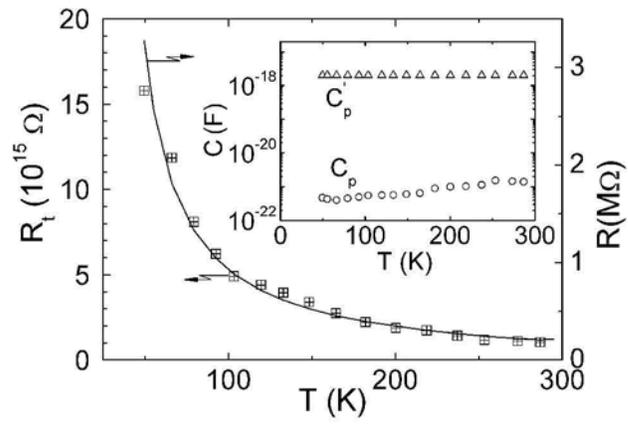